\documentclass[aps,prb,twocolumn,amsmath,amssymb]{revtex4}

\usepackage{graphicx}
\usepackage{wrapfig}
\usepackage{float}
\usepackage{subfig}
\usepackage[section]{placeins}

\usepackage{bm} 
\usepackage{stmaryrd}
\usepackage{txfonts}

\begin{document}

\title{Electronic structure and superconductivity of Europium}

\author{Lane W. Nixon}
\affiliation{ Department of Computational and Data Sciences, George
Mason University, Fairfax, VA 22030}
\author{D. A. Papaconstantopoulos }
\affiliation{ Department of Computational and Data Sciences, George
Mason University, Fairfax, VA 22030}

\begin{abstract}

        We have calculated the electronic structure of Eu for the bcc, hcp, and fcc crystal structures for
volumes near equilibrium up to a calculated 90 GPa pressure using the augmented-plane wave method in the local-density approximation.  
The frozen-core approximation was used with a semi-empirical shift of the f-states energies in the radial Schr$\ddot{o}$dinger 
equation to move the occupied 4f valence states below the $\Gamma_1$ energy and into the core.  This shift of the highly localized 
f-states yields the correct europium phase ordering with lattice parameters and bulk moduli in good agreement with experimental data.  
The calculated superconductivity properties under pressure for the $\it bcc$ and $\it hcp$ structures are also found to agree with 
and follow a $T_c$ trend similar to recent measurement by Debessai et al.~\cite{debessai09}

\end{abstract}

\date{\today}

\maketitle

\section{INTRODUCTION}

	A recent paper by Debessai $\textit{et al.}~\cite{debessai09}$ reports on membrane-driven diamond-anvil experiments on 
europium showing the onset and linear increase of superconductivity under hydrostatic pressure.  A transition temperature near 1.8 K 
was observed near 80 GPa, increasing to 2.75 at 142 GPa.  Previous x-ray diffraction experiments by Takemura and 
Syassen~\cite{takemura85} note the bcc, hcp, hcp-like crystal structures, with transitions occurring at 12.5 and 18 GPa, respectively.  

	Several theoretical works dating back to the mid 1960s have focused on the electronic structure of the equilibrium bcc 
structure.  Freeman and Dimmock reported the first band structure calculations using the non-self-consistent augmented plane-waves (APW) 
method.~\cite{freeman66}  Andersen and Loucks soon followed with relativistic APW (RAPW) method which also was not self 
consistent and used the same exchange potential parameter $\alpha$ = 1.0.~\cite{andersen68}  Kobayasi $\textit{et al.}$ performed similar, 
non-self-consistent calculations with the KKR method with $\alpha$ = 1.0, 0.9, 0.8 and 0.67.~\cite{kobayasi76}
Matsumoto $\textit{et al.}$ related the shape of the Fermi surfaces to helical spin ordering then reported the 
first self-consistent KKR results for $\alpha$ = 1.0 and 0.67, placing the occupied 4f bands near those from 
experiment.~\cite{matsumoto80,matsumoto81}.  Matsumoto $\textit{et al.}$ then reported self-consistent paramagnetic band structure using the 
X$\alpha$-KKR method.~\cite{matsumoto83}  Later Min $\textit{et al.}$ performed self-consistent semi-relativistic and 
fully-relativistic linearized muffin-tin orbital (LMTO) total energy calculations within the local 
density approximation (LDA) for ambient and high pressure bcc Europium.~\cite{min86}  Recently Turek $\textit{et al.}$ calculated the 
exchange parameters, corresponding to magnon spectrum, and N$\dot{e}$el temperature from a real-space perturbation approach derived 
from the tight-binding linear muffin-tin (TB-LMTO) method.~\cite{turek03}  And Kunes and Laskowski calculated the magnetic ground 
state and Fermi surface of bcc Europium using the linearized augmented-plane-waves (LAPW) method with LDA+U to determine spin spiral 
states.~\cite{kunes04}

This work reports the results of paramagnetic self-consistent, scalar-relativistic, frozen-core augmented plane-wave (APW) 
calculations of the total energies and electronic structures of $6s^2$ valence bcc, fcc, and hcp europium for the 
Hedin-Lundqvist~\cite{hedin71} 
local-density approximation (LDA) method with a semi-empirical shift of the f-state energies.

\section{TOTAL ENERGIES}

	In the spirit of previous bcc europium spin-polarized calculations that split localized f-states near $\epsilon_F$ to 
determine structural and magnetic properties in bcc europium (see for example Refs.~\onlinecite{min86,turek03}), self-consistent 
scalar-relativistic frozen core APW total energy calculations were conducted for the bcc, fcc, and hcp structures with a volume-based 
empirical shift of the f-states to determine the ground state and superconductivity properties as functions of pressure.  The 
Brillouin-zone summations over the valence states used 285 bcc, 45 hcp, and 505 fcc k-points in the irreducible portion of the 
Brillouin-zone and a broadening temperature equivalent to 2 mRy.  To simplify the volume-based shift procedure, the hcp structure was 
only considered as ideal with c/a ratio $\surd{\frac{8}{3}}$.  Experiment however shows a decrease in the c/a ratio from 1.60 at 12.5 
GPa to near 1.56 at 18 GPa.~\cite{takemura85}

\begin{figure}[!hb]
\centering
\vspace{-10pt}
\includegraphics[width=0.48\textwidth]{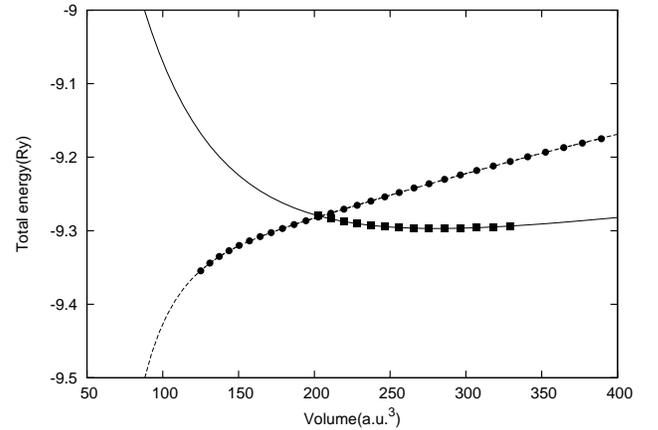}
\vspace{-15pt}
\caption{Europium bcc shifted (solid-line, $\blacksquare$) and unshifted (dashed-line, $\medbullet$) f-states 
total energy showing no minimum.}
\label{fig:comp_energyvolume}
\vspace{-11pt}
\end{figure}

	In a previous paper on lanthanum Ref.~\onlinecite{nixon08} we were able to accurately determine ground state and 
superconductivity properties because the unoccupied 4f-bands are localized well above the Fermi level.  Calculating similar 
properties for europium is problematic because common approximations to density functional theory (DFT) such as the local density 
approximation (LDA) as shown in Fig.~\ref{fig:comp_energyvolume} fail to reliably capture ground state properties due to the 
highly localized 4f-states below the Fermi level.~\cite{turek03}  Figure~\ref{fig:comp_energyvolume} also shows the bcc LDA total 
energies in which we have applied an empirical shift to the f-states as described in detail below.  No energy minimum was 
found for the unshifted f-states calculation which decrease monotonically in energy with volume.

	Since we are most concerned with paramagnetic bulk properties and a reliable ground state, the occupied 4f-states near the Fermi level 
were placed below the $\Gamma_1$ state and treated as part of the core.  The remaining two 6p valence electrons per atom were treated as 
scalar-relativistic bands.  The volume-based shift energy of the f-states was determined in the bcc, fcc, and hcp structures at volume $vol_i$ 
with,

\begin{equation}
E_{f-shift,~vol=i} = \frac{C_{crystal}}{{((\frac{vol_i}{vol_{ref}})^{1/2}{2/3}^{1/4})}}	~,
\label{eqn:shift_eqn}
\end{equation}

\noindent where $C_{crystal}$ was the shift coefficient for the bcc, fcc or hcp structure, $vol_i$ was the volume for the shifted calculation, and 
$vol_{ref}$ was the reference volume.  All structures used the same reference volume of 319.20 $au^3$ which is the volume for the experimental bcc 
lattice a~=~8.61 au.  At this volume the bcc shift coefficient $C_{bcc}$ was 0.6897 Ry while the fcc and hcp shift coefficients, $C_{fcc}$ and 
$C_{hcp}$ respectively, were both $(2/3)^{1/2}C_{bcc}$ which was equivalent to 0.5632 Ry.  The 4f-energies were shifted in the frozen-core self-consistent iterations by adding the shift 
energy ($E_{f-shift,~vol=i}$) to the f-state energies on the right side of the radial Schr$\ddot{o}$dinger equation: 

\begin{equation}
-\frac{1}{r^2}\frac{d}{dr}(r^2\frac{du_l}{dr})+(\frac{l(l+1)}{r^2}+V_n)u_l=Eu_l	~,
\label{eqn:rad_schrodinger}
\end{equation}

\noindent where for $l~=~3, E~=~E~+~E_{f-shift,~vol=i}$.  The bcc shifts yield the energy minimum as shown in Fig.~\ref{fig:comp_energyvolume} 
as the solid-line with black squares.

\begin{figure}[!ht]
\centering
\vspace{-6pt}
\includegraphics[width=0.48\textwidth]{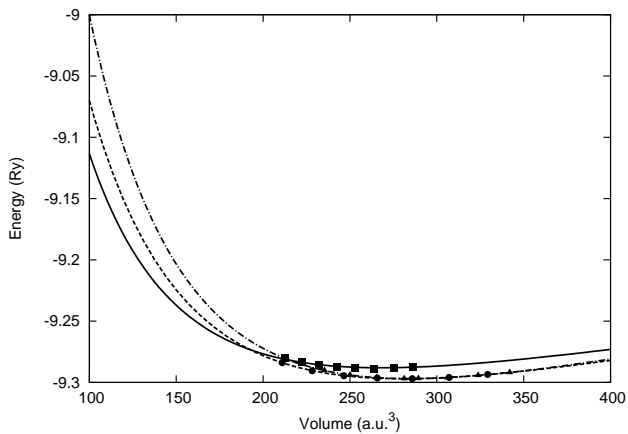}
\caption{Europium hcp (solid-line, $\blacksquare$), bcc (dashed-line, $\medbullet$) and fcc (dash-dot-line, $\blacktriangle$) 
total energies relative to fourth-order Birch fit to bcc structure.}
\label{fig:energyvolume}
\vspace{-6pt}
\end{figure}

	In Fig.~\ref{fig:energyvolume} we show the total energies for europium hcp, bcc, and fcc using the prescribed shifted f-states as a 
function of volume.  The symbols represent the calculated values and the lines are from the $4^{th}$ order Birch fit.  The bcc ground 
state is correctly determined with an equilibrium lattice constant of 8.25 bohr which is about 4 \% below experiment (typical of 
LDA calculations) and a calculated bulk modulus of 14.8 GPa that is about 1 \% greater than experiment.~\cite{nereson64,kittel05}  
The bcc and fcc total energies are very close with bcc equilibrium about 0.3 mRy below the fcc.  

\begin{figure}[!hb]
\centering
\vspace{-6pt}
\includegraphics[width=0.48\textwidth]{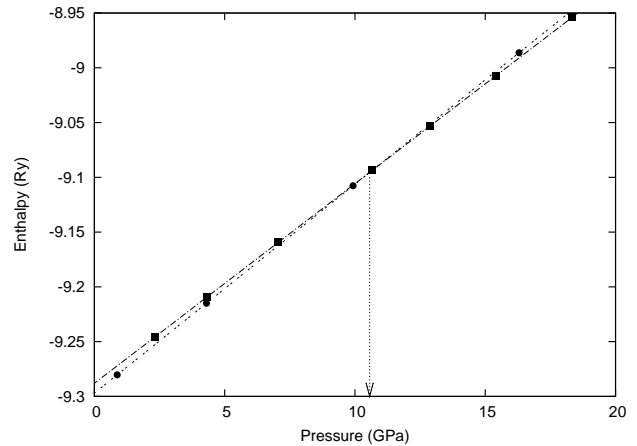}
\caption{Europium hcp (solid-line, $\blacksquare$) and bcc (dashed-line, $\medbullet$) enthalpies.}
\label{fig:enthalpies}
\vspace{-6pt}
\end{figure}

	Since our DFT calculations use temperature T~=0~K, the enthalpy is equal to the Gibbs free energy 
G=E~+pV~-TS such that

\begin{equation}
\vspace{-4pt}
H(P) = E(P)+PV ~,
\label{eqn:enthalpy}
\end{equation}

\begin{figure}[!ht]
\centering
\vspace{6pt}
\includegraphics[width=0.48\textwidth]{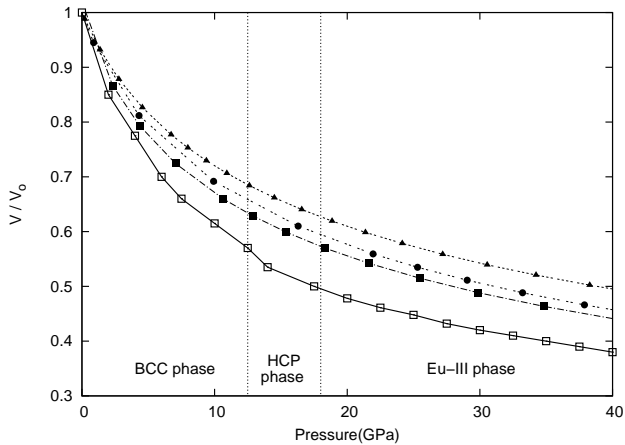}
\caption{Europium hcp (solid-line, $\blacksquare$), bcc (dashed-line, $\medbullet$) and fcc (dash-dot-line, $\blacktriangle$) 
volume ratios as functions of pressure.  Experimental data (solid-line, $\boxempty$) are from Grosshans $\textit{et al.}$ 
(Ref~\protect\onlinecite{grosshans92}).}
\label{fig:VP}
\vspace{6pt}
\end{figure}

\noindent The calculated enthalpies are shown in Fig.~\ref{fig:enthalpies} as functions of pressure for the bcc (circles) and hcp 
(squares) structures.  The transition pressure occurs where the H(P) curves for the bcc and hcp cross near 10.6 GPa which is 15.2~\% 
below the transition observerd at 12.5 GPa in experiment.~\cite{takemura85}  Figure~\ref{fig:VP} compares the calculated volume 
ratios of the hcp, bcc, and fcc as functions of pressure with the ratios from experiment.~\cite{grosshans92}  Though the calculated 
volume ratios as a function of pressure follow a trend similar to experiment they are slightly larger.

	Our bcc results compare well with previous linearized muffin-tin orbital, calculations of Min $\textit{et 
al.}$~\cite{min86}  Their calculations with the paramagnetic 4f electrons as valence states severly overestimate the bonding 
of the localized 4f band, producing a small equilibrium lattice constant 7.61 au and a large bulk modulus 45 GPa.  Their ferromagnetic 
calculations split the 4f states and generated a lattice constant of 8.24 au and bulk modulus of 25 GPa.  
Placing the 4f-electrons in the core significantly improved the bcc equilibrium lattice a~=~8.73 au, which is about 1.4~\% above the experimental 
lattice of 8.61 au for both their paramagnetic and ferromagnetic calculations.  However their paramagnetic and ferromagnetic bulk moduli, which 
are 20 GPa and 19 GPa respectively, are about 33~\% larger than the 15 GPa found in experiment.  Our empirical-shift model achieves bcc lattice 
constants similar to those from the ferromagnetic calculations by Min $\textit{et. al}$ but we also find a bcc equilibrium bulk modulus of 
14.8 GPa that is very close to experiment.

	The results from our first-principles calculations became input to the rigid-muffin-tin approximation (RMTA) theory developed 
by Gaspari and Gyorffy to yield the high-pressure superconducting behavior of bcc and hcp europium, providing additional insight into 
the high pressure $T_c$ trend of Eu-III.~\cite{gaspari72,butler76}

\section{DENSITY OF STATES}

\begin{figure}[!ht]
\centering
\vspace{6pt}
\includegraphics[width=0.48\textwidth]{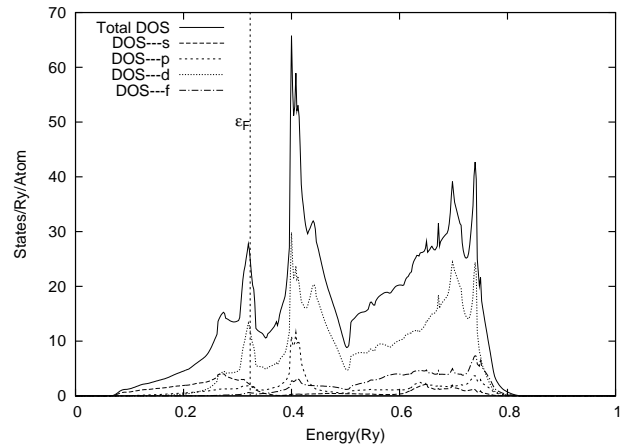}
\caption{Europium shifted f-states bcc equilibruim (a=8.25 a.u.) total and decomposed density of states.}
\label{fig:eu_all_DOS}
\vspace{6pt}
\end{figure}

\begin{figure}[!ht]
\vspace{6pt}
\centering
\includegraphics[width=0.48\textwidth]{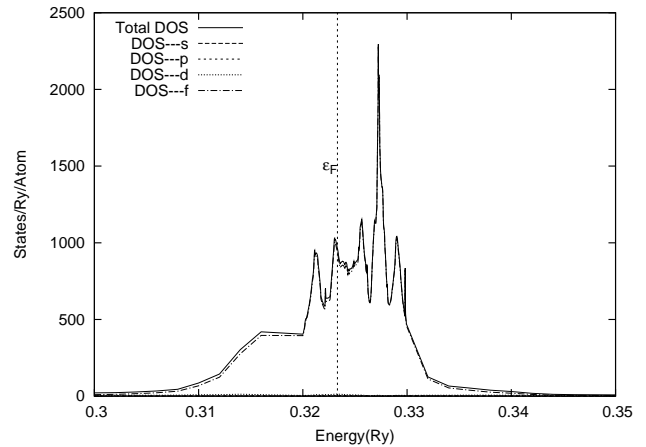}
\caption{Europium unshifted f-states bcc equilibrium (a=8.25 a.u.) total and decomposed density of states at a narrow range near the Fermi 
level.}
\label{fig:eu_bcc_fstate_dos}
\vspace{6pt}
\end{figure}

	The density of states, N($\epsilon$), were calculated by the tetrahedron method~\cite{lehmann72} using the converged 
self-consistent frozen-core bands.  The bcc and fcc N($\epsilon$) were determined using 285 and 505 fcc k-points respectively.  The 
hcp N($\epsilon$) was calculated from Gillan's method~\cite{gillan89} as described in Ref.~\onlinecite{nixon08} but with a 
larger Fermi temperature of 
0.02 Ry to provide additional weight to the broadening function within the k-point mesh.  The sum of the product of energy 
eigenvalues and matrix elements using 45 k-points in the irreducible Brillouin zone yielded the weighted hcp density of states.  
Figures~\ref{fig:eu_all_DOS} and~\ref{fig:eu_bcc_fstate_dos} show the calculated bcc density of states for the shifted and unshifted 
4f-energies respectively at equilibium (a~=~8.25 a.u.).

	The DOS that corresponds to the case of the shifted f-states has the Fermi level, $\epsilon_F$, located to the right of a peak 
near 0.3 Ry and has a large d-like contribution.  The DOS of the unshifted f-states for a narrow region near $\epsilon_F$ are shown 
in Fig.~\ref{fig:eu_bcc_fstate_dos}.  The occupied 4f-states are located below $\epsilon_F$ and the unoccupied 4f-states found above 
$\epsilon_F$.  The unshifted f-states total DOS at $\epsilon_F$, which is composed almost entirely of f-like states, is about 100 
times greater than the shifted N($\epsilon_F$).

\begin{figure}[!ht]
\centering
\subfloat[\bf{hcp $N_l(\epsilon_F)/N_t(\epsilon_F)$}]{\label{fig:eu_hcp_DOS_decomp_full}\includegraphics[width=0.49\textwidth]{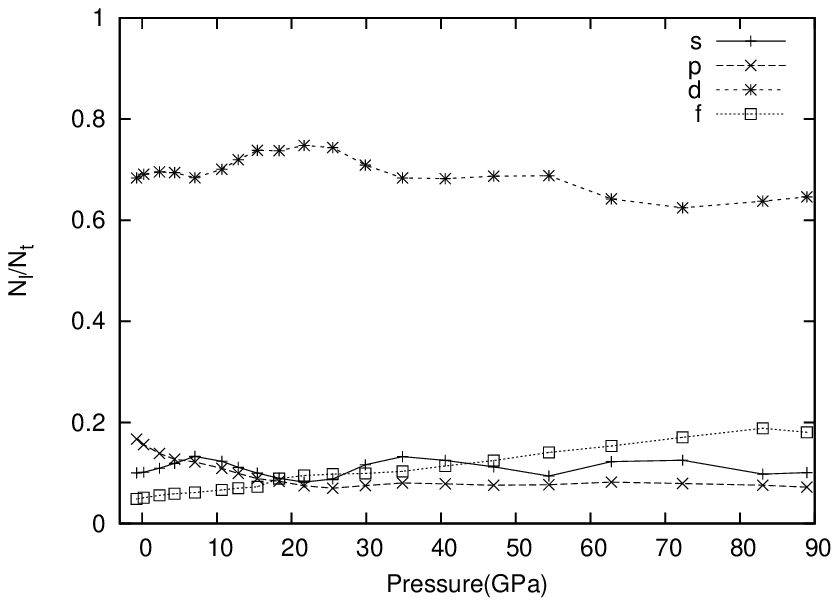}}
\vspace{8pt}
\subfloat[\bf{bcc $N_l(\epsilon_F)/N_t(\epsilon_F)$}]{\label{fig:eu_bcc_DOS_decomp_full}\includegraphics[width=0.49\textwidth]{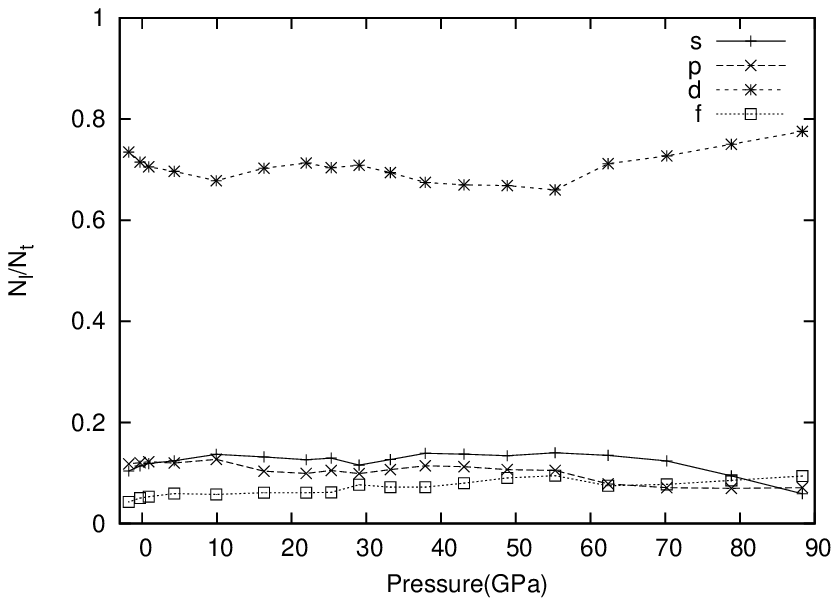}}
\vspace{8pt}
\subfloat[\bf{fcc $N_l(\epsilon_F)/N_t(\epsilon_F)$}]{\label{fig:eu_fcc_DOS_decomp_full}\includegraphics[width=0.49\textwidth]{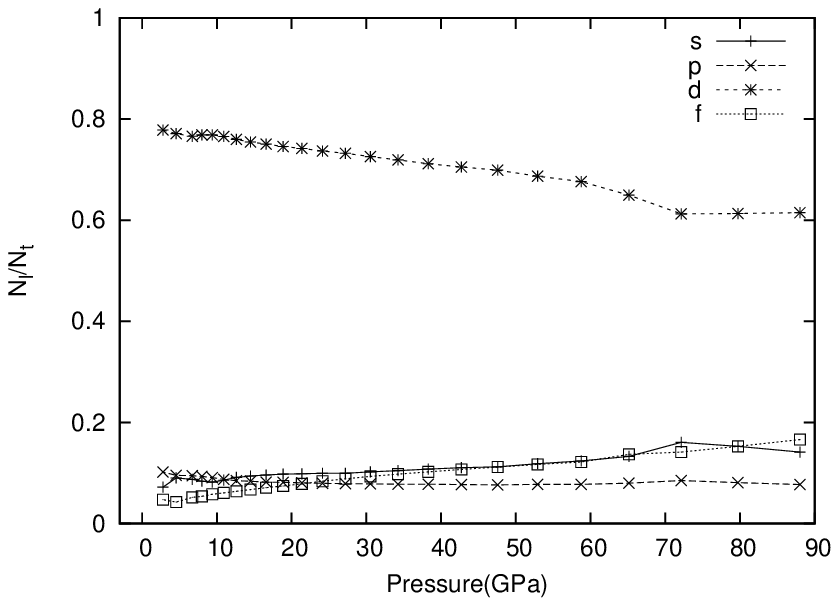}}
\vspace{8pt}
\caption{Europium (a) hcp, (b) bcc, and (c) fcc angular momentum decomposed DOS divided by total DOS at $\epsilon_F$.}
\label{fig:DOS_decomp_all}
\end{figure}

        The significant influence of the d-like states at $\epsilon_F$ is viewed best in Fig.~\ref{fig:DOS_decomp_all}, where the ratio 
N$(\epsilon_F)_l/N(\epsilon_F)$ is displayed as a function of pressure for the hcp (a), bcc (b), and fcc (c) structures.  The 
dominant d-like states remain fairly constant showing only moderate changes with pressure.  The hcp d-like states at $\epsilon_F$ 
increase to almost 0.8 near 24 GPa, drop to a low of about 0.61 near 70 GPa then increase at higher pressures.  The bcc d-like states 
at $\epsilon_F$ decrease from 0.73 at equilibrium to 0.68 at near 12.5 GPa.  Between 12.5 GPa and 58 GPa the bcc d-like states 
increase back to 0.73 then return to about 0.68 near 58 GPa.  At higher pressures the bcc d-like states increase reaching a maximum 
of almost 0.78 near 90 GPa.  And the fcc d-like states at $\epsilon_F$ decrease almost monatonically from equilibrium pressure to 
about 0.60 near 70 GPa.  The fcc d-like states show only minor increases at higher pressures.

	The hcp f-like states at $\epsilon_F$ increase monatonically with pressure with relatively no change between 25 and 35 GPa and 
a slight decrease after about 85 GPa.  At equilibrium pressure the hcp f-like states are about 0.05 and reach a maximum of almost 0.2 
near 85 GPa.  The bcc f-like states at $\epsilon_F$ increases from about 0.05 at equilibrium to almost 0.1 near 58 GPa, decreases to 
about 0.07 above 60 GPa, and increases back to almost 0.1 near 90 GPa.  The fcc f-like states at $\epsilon_F$ increase monatonically 
with pressure from about 0.035 near 4 GPa to almost 0.17 near 90 GPa.

        The hcp s-like density of states at $\epsilon_F$ oscillate with pressure from a high of almost 0.16 to a low near 0.1.  The 
bcc s-like states at $\epsilon_F$ show only minor changes from equilibrium pressure to about 62 GPa.  Afterwards they decrease to 
almost 0.06 near 90 GPa.  The fcc s-like states increase monatonically from almost 0.06 near 4 GPa to about 0.15 slightly with 
pressure to equilibrium where is remains relatively constant with pressure. 

	The hcp p-like N($\epsilon_F$) ratio decreases from about 0.18 at equilibrium to about 0.09 at 20 GPa where it shows no change with 
pressure.  The bcc p-like N($\epsilon_F$) decreases slowly from near 0.12 at equilibrium to about 0.06 near 62 GPa, remaining constant with 
pressure.  And the fcc p-like N($\epsilon$) remains constant with pressure from about 0.08 near 4 GPa.

	Overall the $N_l(\epsilon_F)/N_t(\epsilon_F)$ ratios vary slowly with pressure.

\vspace{6pt}
\subsection{Band Structure}
\vspace{6pt}

\begin{figure*}[!ht]
\centering
\vspace{-6pt}
\subfloat[Shifted 4f-states]{\label{fig:band_sh}\includegraphics[width=0.49\textwidth]{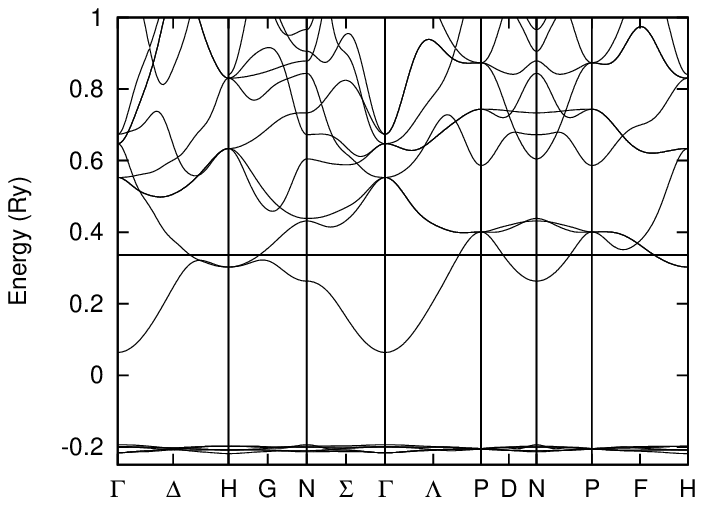}}
\subfloat[Unshifted 4f-states]{\label{fig:band_un}\includegraphics[width=0.49\textwidth]{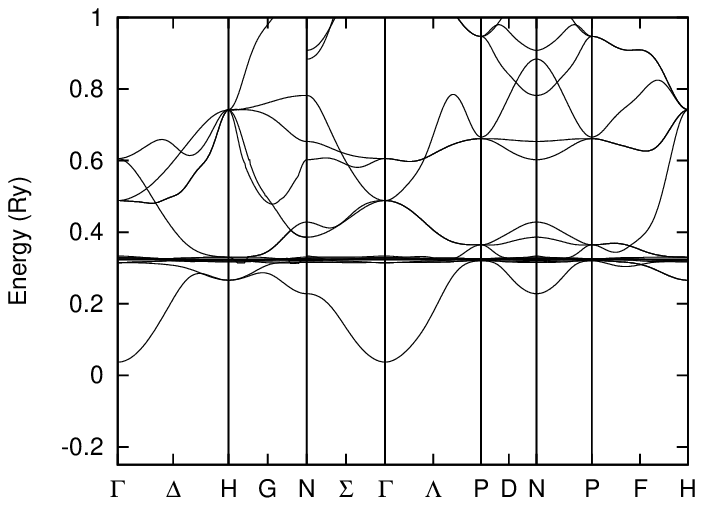}}
\caption{Europium equilibrium bcc shifted and unshifted f-states band structures. }
\label{fig:eu_bands}
\vspace{-6pt}
\end{figure*}

	Figure~\ref{fig:eu_bands} shows the band structures of bcc equilibrium respectively with the shift of the energy of the 4f-states 
(Fig.~\ref{fig:band_sh}) and without the shift (Fig.~\ref{fig:band_un}).  The shifted 4f-states are below 
the $\Gamma_1$ energy and highly localized within a energy block less than 0.13 Ry.  The unshifted 4f-states are still highly 
localized but entangle themselves near the Fermi level, making it impossible to determine the minimum energy for the structure.

	Our empirical-shift model explicitly omits hybridization of the occupied 4f levels by placing them below the $\Gamma_1$ level, they 
contribute little to chemical bonding.~\cite{richter98}  Spin-polarized calculations by Kunes and Laskowski, Min $\textit{et al.}$, 
and others that segregate the occupied and unoccupied 4f states respectively below and above $\epsilon_F$ report negligible mixing of 
the occupied 4f states.~\cite{kunes04,min86}  Recent photoemission studies by Wang $\textit{et al.}$ reports the highly localized 4f 
core level at about 0.154 Ry below the Fermi level.  Replicating the 4f core location with such proximity to $\epsilon_F$ would be 
difficult even with the more sophisticated LDA+U or the self-interaction corrected (SIC) LSDA approaches.

\section{SUPERCONDUCTIVITY}

	From our densities of states results at the Fermi level we determine, from first-principles, 
the electron-phonon coupling constant $\lambda$ and the superconducting transition temperature $T_c$.  Using 
McMillan's strong coupling theory,~\cite{butler76} we take the electron-phonon coupling constant by the following formula: 

\begin{equation}
\vspace{-8pt}
\lambda = \frac{\eta}{M <\omega^2>} ~ ,
\label{eqn:lambda}
\end{equation}

\noindent where M is the atomic mass, $\eta$ is the Hopfield parameter~\cite{hopfield69} calculated with the 
RMTA,~\cite{gaspari72,butler76} N($\epsilon_F$) is the total density of states at the Fermi level, $\epsilon_F$, $<I^2>$ is the 
square of the electron-ion matrix element at $\epsilon_F$, and $<\omega^2>$is the square of the average phonon frequency.  The 
Hopfield parameter is determined by the quantity 

\begin{equation}
\vspace{-8pt}
\eta = N(\epsilon_F)<I^2> ~ ,
\label{eqn:eta}
\end{equation}

\noindent using the RMTA. The square matrix element $<I^2>$, which is derived from multiple-scattering theory and is also determined 
by the RMTA proposed by Gaspari and Gyorffy, is defined by the formula 

\begin{equation}
\vspace{-8pt}
<I^2> = \frac{\epsilon_F}{\pi^2N^2(\epsilon_F)}\Sigma_l
\frac{2(l+1)sin^2(\delta_{l+1} - \delta_l)N_lN_{l+1}}
{N_{l}^{(1)}N_{l+1}^{(1)}} ~ ,
\label{eqn:matrix_elem}
\end{equation}

\noindent where $\delta_l$ are scattering phase shifts, $N_l(\epsilon_F)$ is the ${l}^{th}$ component of the DOS per spin, and 
$N_l^{(1)}$ is the free-scatter DOS,

\begin{equation}
\vspace{-8pt}
N^{(1)}_l=\frac{\surd{\epsilon_F}}{\pi}(2l+1)\int^{R_s}_0u^2_l(r,\epsilon_F)r^2dr~.
\label{eqn:scatters}
\end{equation}

The scattering phase shifts $\delta_l$ were calculated from the radial wave functions $u_l$, spherical Bessel functions $j_l$ and 
Neumann functions $n_l$ at the muffin-tin radius $R_s$ from the formula

\begin{equation}
\vspace{-8pt}
\frac{u`_l(r,\epsilon_F)}{u_l(r,\epsilon_F)}=\frac{j`_l(kr)-n`_l(kr)tan~\delta_l}{j_l(kr)-n_l(kr)tan~\delta_l}|_{r=R_s}~,
\label{eqn:deltas}
\end{equation}

\noindent where the shift energy ($E_{f-shift,~vol=i}$) was added to the Fermi energy of the $4^{th}$ component l=3 in both 
Eqns.~\ref{eqn:scatters} and~\ref{eqn:deltas}.  Equation~\ref{eqn:matrix_elem} is used with touching MT spheres to minimize errors 
due to excess interstitial volume as demonstrated in Ref.~\onlinecite{lei07}, ~\onlinecite{nixon07}, and~\onlinecite{nixon08}.  The 
APW total energy results were used to calculate the pressure variation of the bulk moduli B.  These values were used to determine the 
pressure variation of the Debye temperature $\theta_D$, which has no experimentally measured bcc equilibrium $\theta_D$ below a 
temperature of 100~K,~\cite{lounasmaa65} using the formula by Moruzzi $\textit{et al.}$:\cite{moruzzi88}

\begin{equation}
\vspace{-8pt}
\theta_D=213.4\surd{\frac{r_oB}{M}}~ ,
\label{eqn:moruzzi}
\end{equation}

\noindent where $B$ is the bulk modulus (in GPa), $r_o$ is the Wigner-Seitz radius in Bohr units, and $M$ is the atomic mass.  The 
approximate average phonon frequency $<\omega^2>$ is determined as

\begin{equation}
\vspace{-8pt}
<\omega^2>=\frac{1}{2}\theta^2_D~,
\label{eqn:omega}
\end{equation}

\noindent and is used both in the denominator of Eq.~\ref{eqn:lambda} and in the prefactor of the McMillan equation for 
the determination of the superconductivity transision temperature $T_c$,~\cite{mcmillan68}

\begin{equation}
\vspace{-8pt}
T_c = \frac{<\omega>}{1.2}exp[{\frac{-1.04(1+\lambda)}{\lambda-\mu*(1+0.62\lambda)}}] ~ ,
\label{eqn:tc}
\end{equation}

\noindent where the Coulomb pseudopotential $\mu*$ was fixed at 0.13 for all three structures.

\begin{figure*}[!ht]
\vspace{-10pt}
\centering
\subfloat{\label{fig:eu_bulk}\includegraphics[width=0.48\textwidth]{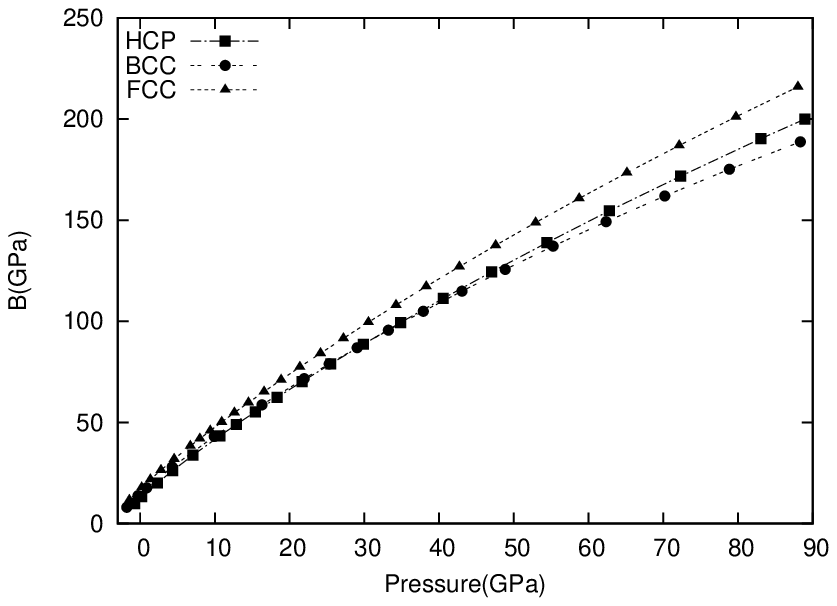}}
\subfloat{\label{fig:eu_omega}\includegraphics[width=0.48\textwidth]{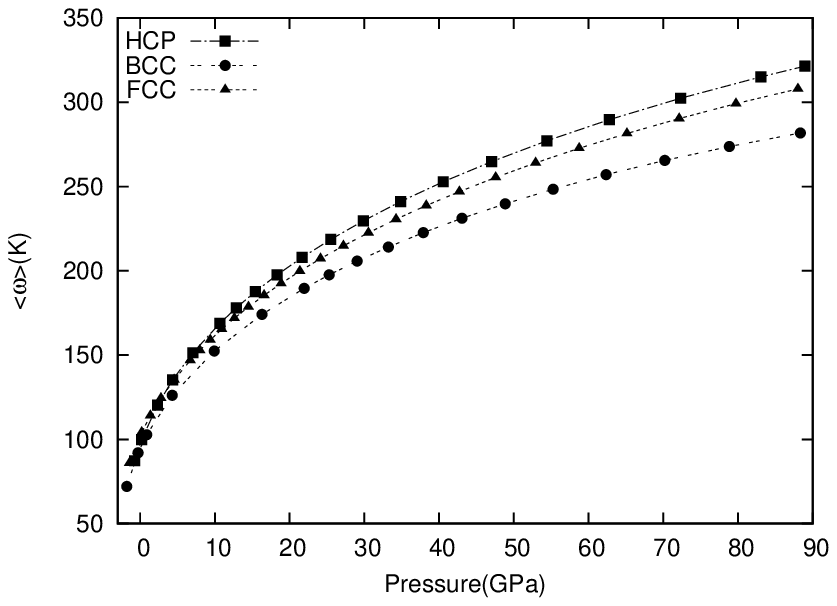}}
\caption{Bulk modulus B and average phonon frequency $<\omega>$ of hcp (solid-line, $\blacksquare$), bcc (dashed-line, $\medbullet$) and fcc 
(dash-dot-line, $\blacktriangle$) as functions of pressure.}
\label{fig:eu_bulk_omega}
\vspace{-10pt}
\end{figure*}

	From our total energy and density of states calculations we determined the input parameters for the $T_c$ calculations for the 
shifted 4f-states for the bcc, hcp, and fcc structures.  The bulk modulus for Eu bcc is shown in Figure~\ref{fig:eu_bulk} as a 
function of pressure.  The average phonon frequencies show a similar trend, as displayed in Figure~\ref{fig:eu_omega}, since they are 
derived from the bulk moduli.  These trends, where B is directly related to $<\omega>$, also agree qualitatively with the 
proportionality relations used for the lighter elements of fcc yttrium,~\cite{lei07} fcc scandium,~\cite{nixon07} and 
lanthanum.~\cite{nixon08}

\begin{figure*}[!ht]
\centering
\subfloat[N($\epsilon_F$)]{\label{fig:eu_nef_spin}\includegraphics[width=0.48\textwidth]{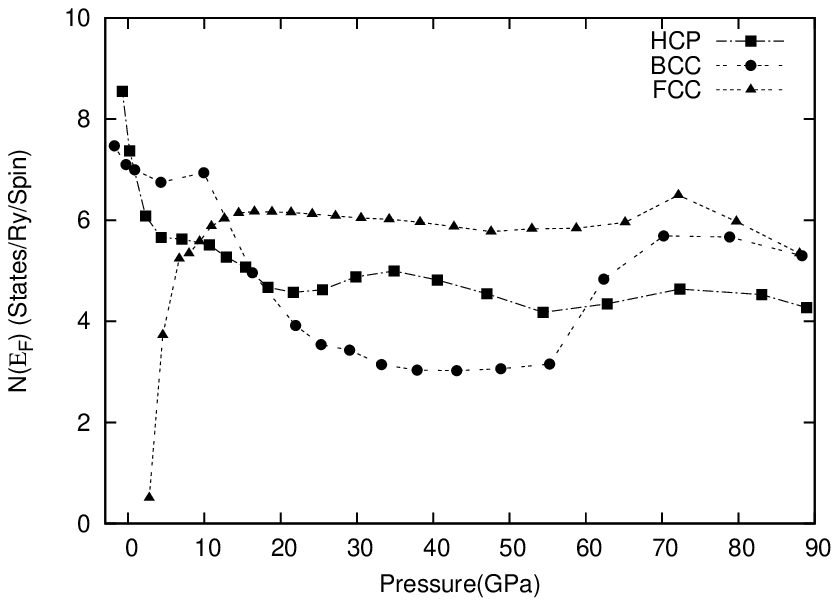}}
\subfloat[$<I^2>$]{\label{fig:eu_elect_ion}\includegraphics[width=0.48\textwidth]{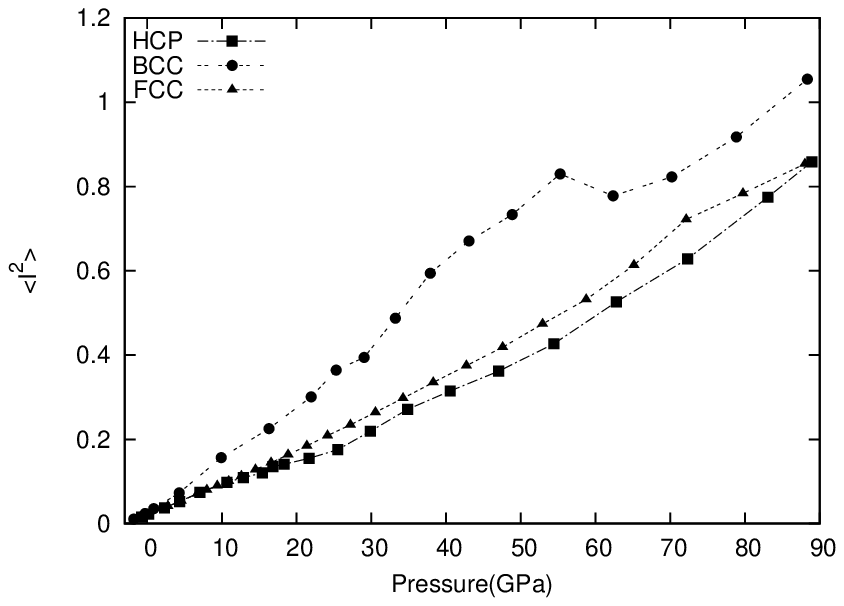}}
\vspace{-10pt} 
\caption{Total DOS per spin at $\epsilon_F$ and electron-ion matrix element $<I^2>$ of hcp (solid-line, $\blacksquare$), bcc (dashed-line, 
$\medbullet$) and fcc (dash-dot-line, $\blacktriangle$) as functions of 
pressure.}
\label{fig:DOS_elect_ion}
\vspace{-6pt}
\end{figure*}

\begin{figure*}[!ht] 
\vspace{-6pt} 
\centering 
\subfloat{\label{fig:eu_bcc_comp_phase}\includegraphics[width=0.48\textwidth]{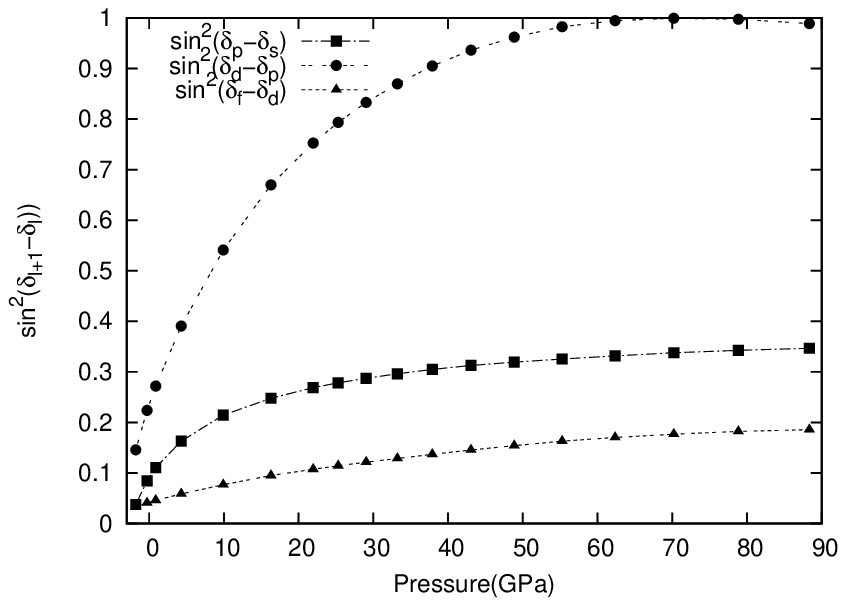}}
\subfloat{\label{fig:eu_bcc_comp_eta}\includegraphics[width=0.48\textwidth]{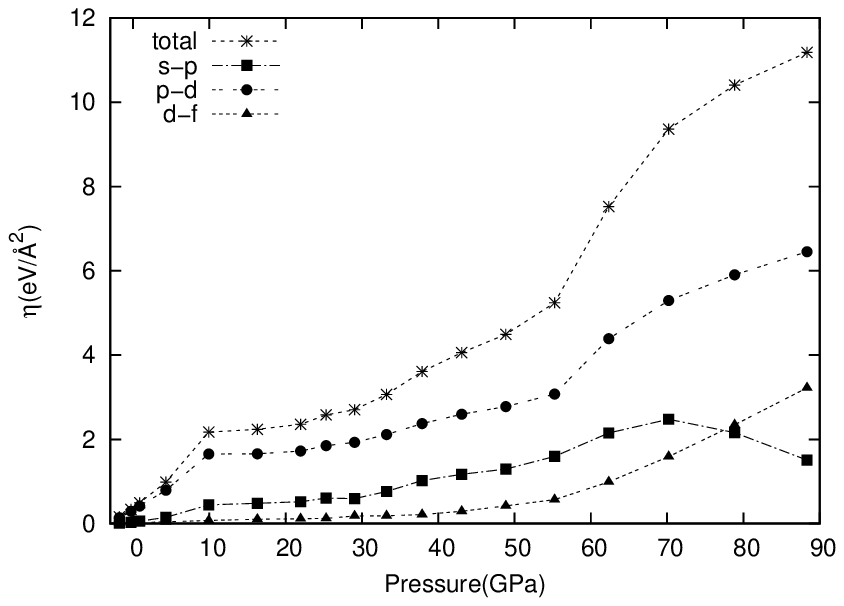}}
\vspace{-10pt} 
\caption{The phase shift factor $sin^2(\delta_{l+1}-\delta_1)$ for $l=0$ (squares), $l=1$ (circles), and $l=2$ 
(triangles) and the total and decomposed Hopfield parameter $\eta$ of bcc europium as functions of pressure.} 
\label{fig:comp_phase_eta} 
\vspace{-10pt} 
\end{figure*}

\begin{figure*}[!ht]
\centering
\subfloat[$\eta$]{\label{fig:eu_eta}\includegraphics[width=0.48\textwidth]{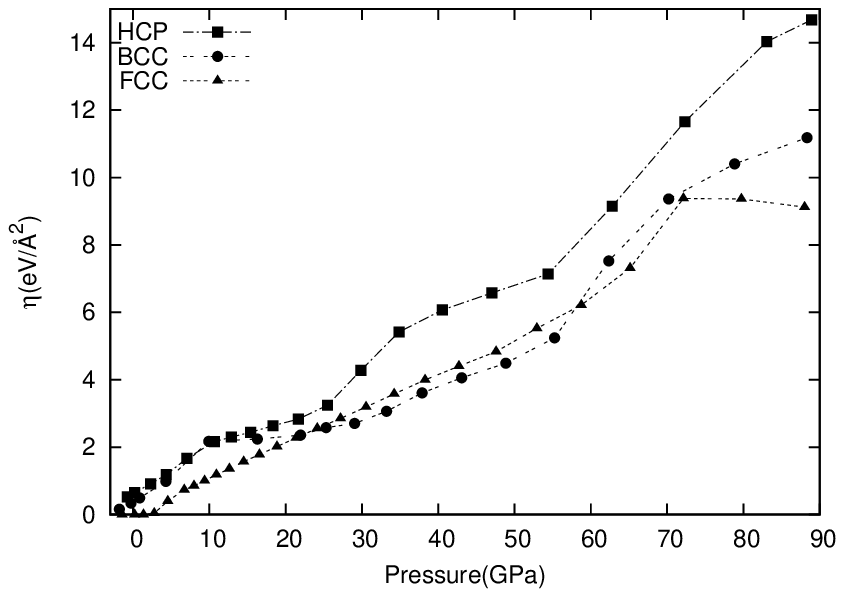}}
\subfloat[$\lambda$]{\label{fig:eu_lambda}\includegraphics[width=0.48\textwidth]{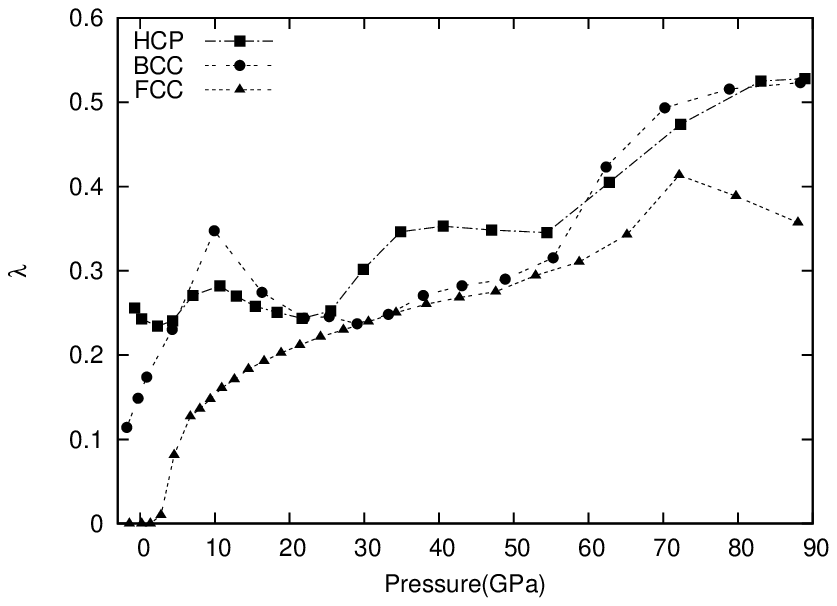}}
\vspace{-10pt} 
\caption{Hopfield parameter $\eta$ and electron-phonon coupling constant
$\lambda$ of hcp (solid-line, $\blacksquare$), bcc (dashed-line, $\medbullet$) and fcc (dash-dot-line, $\blacktriangle$) as functions of 
pressure.}
\label{fig:DOS_eta_lambda}
\vspace{-10pt}
\end{figure*}

	The density of states at the Fermi level N$(\epsilon_F)$, electron-ion matrix element $<I^2>$, bcc phase shift factor, and bcc 
total and decomposed Hopfield parameter $\eta$ as functions of pressure are shown in Figs.~\ref{fig:eu_nef_spin}, 
~\ref{fig:eu_elect_ion},  ~\ref{fig:eu_bcc_comp_phase}, and~\ref{fig:eu_bcc_comp_eta}, respectively.  After about 20 GPa the density 
of states varies slowly with pressure to 90 GPa for the hcp and fcc.  The bcc N($\epsilon_F$) increase rapidly after about 55 GPa 
from near 3.7 states/Ry/spin to almost 5.8 states/Ry/spin near 70 GPa.  The hcp and fcc N($\epsilon_F$) also decrease after 70 GPa.  

\begin{figure*}[!ht]
\centering
\includegraphics[width=0.8\textwidth]{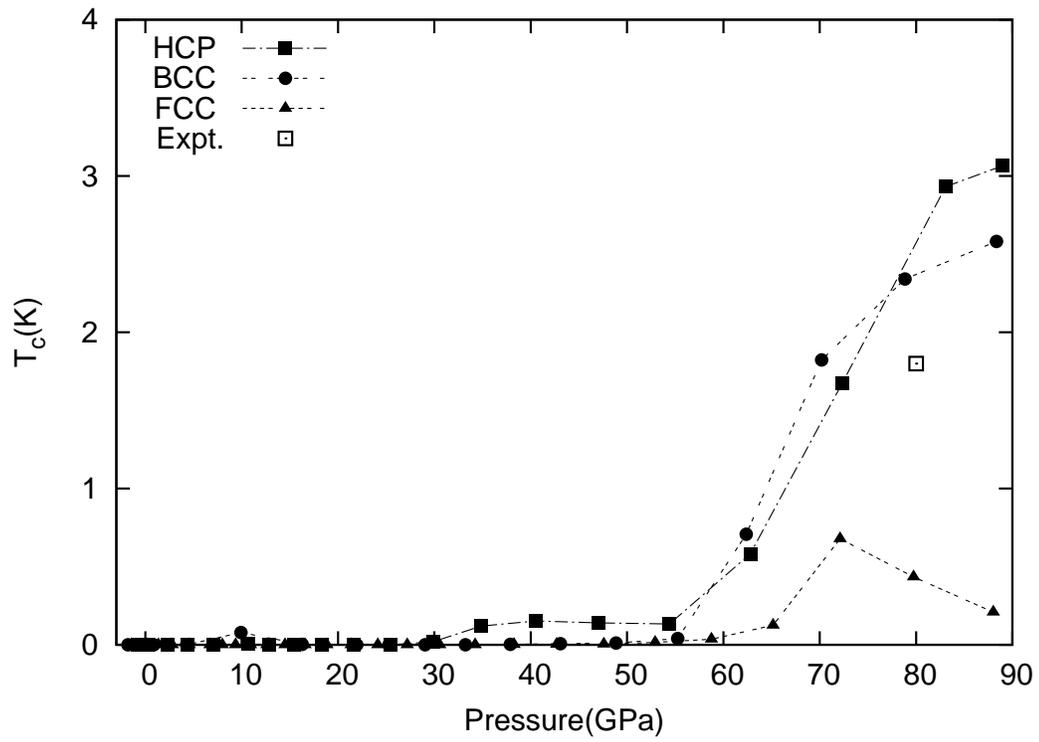}
\caption{Transition temperature $T_c$ of hcp, bcc, and fcc with $\mu*$ = 0.13 as functions of pressure.  Experimental data from 
Debessai $\textit{et al.}$ 
(Ref.~\protect\onlinecite{debessai09}).}
\label{fig:eu_Tc}
\vspace{-8pt}
\end{figure*}

	Although the N($\epsilon_F$) vary slowly with pressure, the electron-ion matrix element $<I^2>$ in Eqn.~\ref{eqn:eta} 
increases rapidly with pressure in all three structures.  The increase of the electron-ion matrix element $<I^2>$ comes primarily 
from the large p-d phase shift factor which is near one at high pressures as shown in Fig.~\ref{fig:eu_bcc_comp_phase} for bcc.  
Figure~\ref{fig:eu_bcc_comp_eta} shows the bcc total and decomposed Hopfield parameter as a function of pressure.  The large increase 
in total $\eta$ with pressure comes from the large p-d $\eta$ contribution.  The hcp and fcc also have large p-d phase factor and 
$\eta$ components that increase the total $\eta$.

	Figures~\ref{fig:eu_eta} and~\ref{fig:eu_lambda} respectively show the total $\eta$ and electron-phonon coupling constant 
$\lambda$ as function of pressure for the hcp, bcc, and fcc.  The increasing $\eta$ values in the numerator of Eqn.~\ref{eqn:lambda} 
along with moderate phonon frequencies in the denominator help explain the increase of the electron-phonon coupling constant $\lambda$ 
with pressure.  Above 55 GPa the bcc and hcp $\lambda$ goes above 0.4 producing an increase in the superconductity transistion 
temperature $T_c$ shown in Fig.~\ref{fig:eu_Tc}.  The calculated bcc and hcp $T_c$ flattens near 80 GPa because the $\lambda$ levels 
off at about 0.53 near the same pressure range.  The fcc shows a peak in $T_c$ near 70 GPa but decreases at higher pressures due to 
the decreasing fcc $\eta$ that yields smaller $\lambda$ with pressure.

\clearpage

\vspace{-10pt}
\section{CONCLUSIONS}

	In conclusion, the electronic structure and superconductivity properties of europium have been 
studied as functions of pressure using the Hedin-Lundqvist LDA form of the exchange-correlation and shifted 4f-band energies within 
the APW method.  In particular, we investigated the total energies for the bcc, fcc, and hcp structures; the density of 
states, for the bcc and hcp structures and the band structure for the 
bcc and hcp; and the superconducting properties for the bcc and hcp structures through high 
pressures, using the RMTA.

	On the basis of our total energy calculations, the bcc lattice has been shown to be the equilibrium structure, while the hcp 
becomes favorable at high pressures.  Our density of states calculations agree well with previous results for 
the bcc structure.  Our calculations of the superconducting properties show a gradual increase of the electron-phonon coupling 
constant to values around 0.5, reproducing the onset of superconductivity at pressures between 60 and 90 GPa.

\vspace{-15pt}	

\section{Acknowledgements}
LWN was supported by a National Science Foundation GK-12 Fellowship, George Mason University, under grant DGE 0638680.  This work was 
also partially supported by NIST grant 70NANB7H6138.

\end{document}